\documentclass{osa-article}

\journal{osajournal}

\articletype{Research Article}

\begin{document}

\title{Flattening laser frequency comb spectra with a high dynamic range, broadband spectral shaper on-a-chip}

\author{Nemanja Jovanovic,\authormark{1,*} Pradip Gatkine,\authormark{1} Boqiang Shen,\authormark{1,2} Maodong Gao,\authormark{1,2}  Nick Cvetojevic,\authormark{3} Katarzyna Ławniczuk,\authormark{4} Ronald Broeke,\authormark{4} Charles Beichman,\authormark{5} Stephanie Leifer,\authormark{6} Jeffery Jewell,\authormark{5} Gautam Vasisht,\authormark{5} and Dimitri Mawet\authormark{1,5}}

\address{\authormark{1}Department of Astronomy, California Institute of Technology, Pasadena, CA, 91125, USA\\
\authormark{2}T. J. Watson Laboratory of Applied Physics, California Institute of Technology, Pasadena, CA, 91125, USA\\
\authormark{3}Université Côte d'Azur, Observatoire de la Côte d'Azur, CNRS, Laboratoire Lagrange, France\\
\authormark{4}Bright Photonics BV, Horsten 1, 5612 AX Eindhoven, The Netherlands\\
\authormark{5} NASA Exoplanet Science Institute, Jet Propulsion Laboratory, California Institute of Technology, 4800 Oak Grove Drive, Pasadena, CA 91109, USA\\
\authormark{6}The Aerospace Corporation, 2310 E. El Segundo Blvd., El Segundo, CA 90245\\}



\email{\authormark{*}nem@caltech.edu} 



\begin{abstract}
Spectral shaping is critical to many fields of science. In astronomy for example, the detection of exoplanets via the Doppler effect hinges on the ability to calibrate a high resolution spectrograph. Laser frequency combs can be used for this, but the wildly varying intensity across the spectrum can make it impossible to optimally utilize the entire comb, leading to a reduced overall precision of calibration. To circumvent this, astronomical applications of laser frequency combs rely on a bulk optic setup which can flatten the output spectrum before sending it to the spectrograph. Such flatteners require complex and expensive optical elements like spatial light modulators and have non-negligible bench top footprints. Here we present an alternative in the form of an all-photonic spectral shaper that can be used to flatten the spectrum of a laser frequency comb. The device consists of a circuit etched into a silicon nitride wafer that supports an arrayed-waveguide grating to disperse the light over hundreds of nanometers in wavelength, followed by Mach-Zehnder interferometers to control the amplitude of each channel, thermo-optic phase modulators to phase the channels and a second arrayed-waveguide grating to recombine the spectrum. The demonstrator device operates from 1400 to 1800 nm (covering the astronomical H band), with twenty 20 nm wide channels. The device allows for nearly 40 dBs of dynamic modulation of the spectrum via the Mach-Zehnders , which is greater than that offered by most spatial light modulators. With a smooth spectrum light source (superluminescent diode), we reduced the static spectral variation to $\sim$3 dB, limited by the properties of the components used in the circuit. On a laser frequency comb which had strong spectral modulations, and some at high spatial frequencies, we nevertheless managed to reduce the modulation to $\sim$5 dBs, sufficient for astronomical applications. The size of the device is of the order of a US quarter, significantly cheaper than their bulk optic counter parts and will be beneficial to any area of science that requires spectral shaping over a broad range, with high dynamic range, including exoplanet detection.   
\end{abstract}

\section{Introduction}

The ability to shape the spectrum of light is critical to many applications including temporal pulse shaping~\cite{kurokawa1999-TSC,Miyamoto2006-WCO,WEINER2011-UOP,Roelens2008-MWS}, gain flattening~\cite{Lewis2000-RNS,Harumoto2002-GFF}, add/drop multiplexing~\cite{baxter2006-HPW} and other applications in telecommunications as well as the targeted excitation of particular molecular species~\cite{moroz2017-LSC}, to name a few. In the field of astronomy, shaping the spectrum of a Laser Frequency Comb (LFC), and more specifically flattening it, is essential to maximize the signal-to-noise ratio of as many comb lines as possible~\cite{probst2013-SFS}. 
By doing so, this leads to an overall improvement in the accuracy of the calibration of the spectrograph. Such calibrations directly impact the ability for high resolution spectrographs to detect exoplanets (planets around stars other than our sun) via the Doppler or radial velocity techniques, which gets more difficult as the planets mass decreases~\cite{Fischer2016-EPR}.    

\begin{figure}[htbp]
\centering\includegraphics[width = \linewidth]{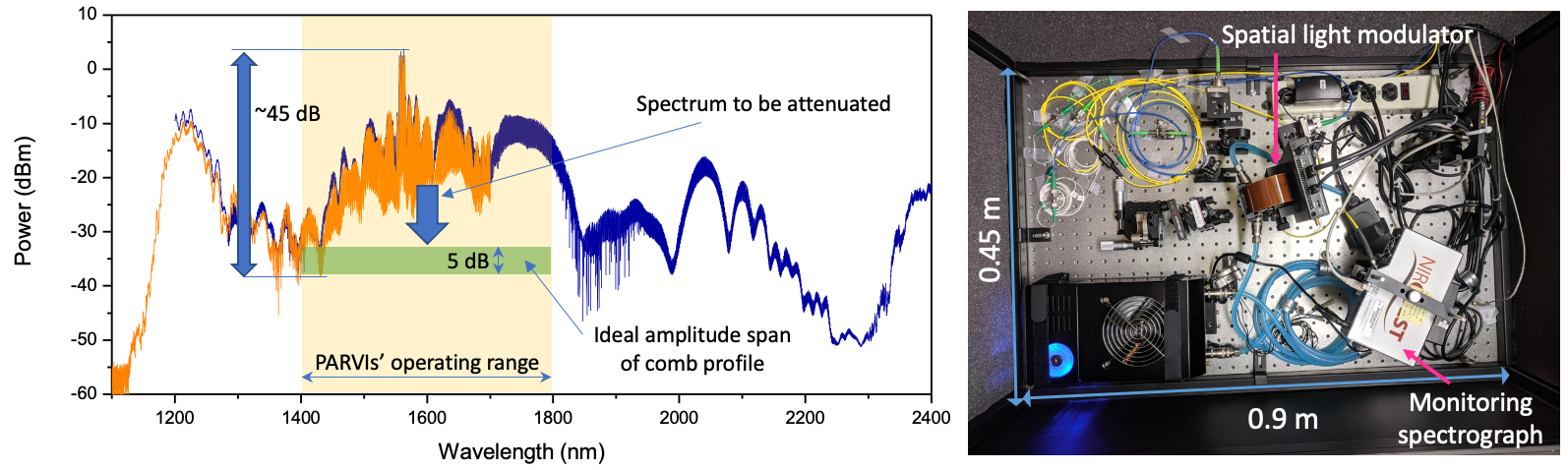}
\caption{(Left) a spectrum of the PARVI LFC showing the strong amplitude modulations across the spectrum. (Right) PARVI's bulk optic flattener as an example. The flattener occupies a volume of $1\times0.5\times0.5$ m$^{3}$. \label{fig:bulkflattener}}
\end{figure}

An example of the issue can be seen in the left panel of Fig.~\ref{fig:bulkflattener}. This shows a trace from the LFC spectrum used with the PARVI spectrograph at Palomar Observatory~\cite{Gibson2019}, which exhibits amplitude differences of up to 45 dB across the 400 nm window, from 1400 to 1800 nm used by the instrument currently. 
To combat this problem, spectral flatteners are typically constructed from bulk optics, where light is dispersed with a grating and a spatial light modulator is used to control the amplitude across the spectrum, before recombining the light into a fiber and sending it to the spectrograph~\cite{probst2013-SFS}. As seen in the right panel of Fig.~\ref{fig:bulkflattener}, these setups can be large (small bench tops) and expensive (several hundred thousand dollars based on two flatteners purchased by this team around the time of this article). Reducing the volume of these setups, and integrating them to maximize stability would be ideal. Given that the final comb spectrum is typically generated inside a waveguide or optical fiber, finding an all photonic solution to flattening the spectrum would be the most elegant way of miniaturizing this necessary component.

In addition, the brightness of the comb lines can vary with time. The evolution of the comb profile of an LFC can vary over a broad range of timescales from milliseconds to hours and is significantly smaller (1 to 10 dB) than the amplitude differences across the spectrum. Unfortunately, any changes to the comb profile can masquerade as erroneous Doppler shifts and needs to be mitigated. These variations necessitate that the flattener be a dynamic element that can track and control the evolution in the spectrum with time. 

To address both of these issues, we have developed a broadband, all-photonic spectrum shaper/flattener (BAPSS) on-a-chip. The device was constructed from optical waveguides on a silicon nitride (SiN) chip and operates from 1400 to 1800 nm (covering the astronomical H band), with 20, 20 nm wide channels. The device was mounted on a PCB to enable electrical control of the active elements and tested in the laboratory. Section~\ref{sec:design} outlines the design and simulated performance of the photonic circuit. Section~\ref{sec:experiments} provides an overview of the experimental procedures while Section~\ref{sec:results} highlights the key results of the test devices. Section~\ref{sec:discussion} concludes the paper by providing a discussion about the results and their implications.

\section{Circuit design and simulated performance}\label{sec:design}

\begin{figure}[htbp]
\centering\includegraphics[width = \linewidth]{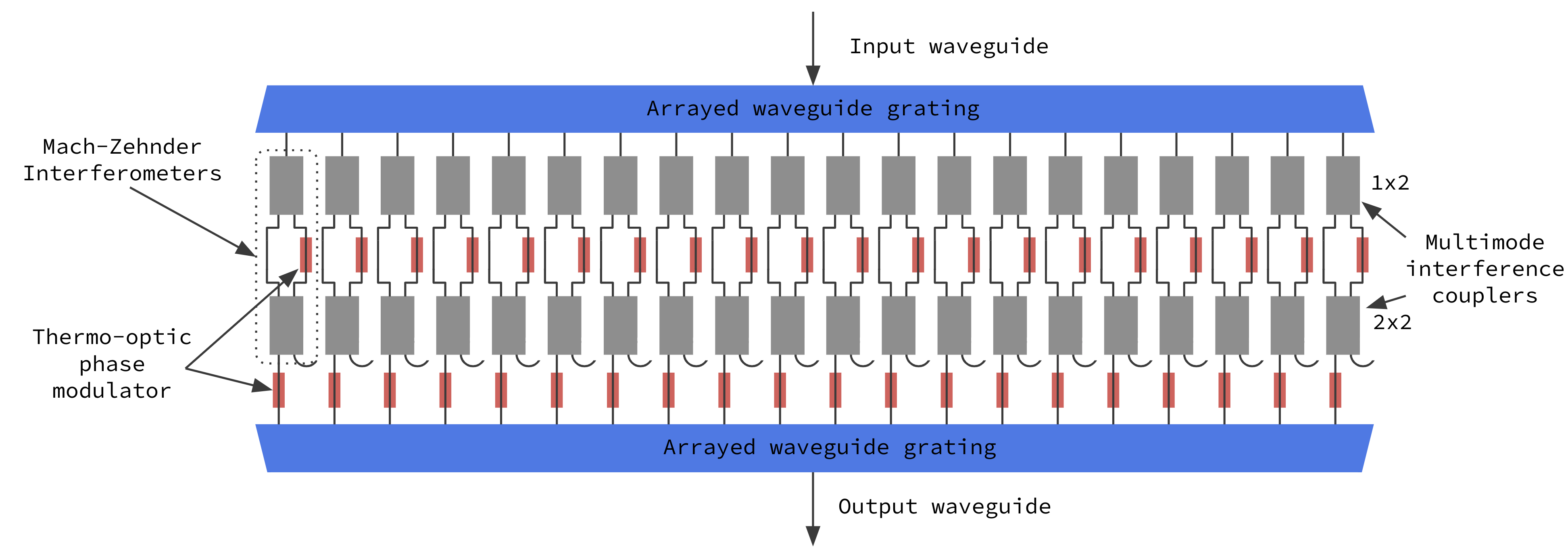}
\caption{Schematic layout of the circuit architecture used for the photonic spectral shaper. \label{fig:concept}}
\end{figure}

The architecture of the circuit we choose for this application is shown in Fig.~\ref{fig:concept} and consists of the following elements:
\begin{itemize}
    \item An arrayed waveguide grating (AWG) to first disperse the light into 20 discrete, narrow band (20 nm) channels
    \item Mach-Zehnder Interferometers (MZIs) in each AWG output, with adjustable path lengths implemented on one arm via Thermo-Optic Phase Modulators (TOPMs). The transmitted power through the MZI can be adjusted by varying the phase of one arm.  
    \item TOPMs following each MZI to adjust the path length of a given spectral channel with respect to its neighbor, in order to rephase the spectrum when its recombined and 
    \item Another AWG with the same properties as the first to recombine the channels into a single spectrum. 
\end{itemize}
This architecture is based on the design proposed in~\cite{Takiguchi-FPW2004} (uses the same components in the same order) and is similar to~\cite{Miyamoto2006-WCO} (which instead uses a reflective design and LiNbO$_{3}$ phase modulators) for optical synthesis applications, but is optimized in a different way to meet the needs of flattening the light from an LFC. Specifically, the device presented here is optimized to operate over a broad wavelength range (400 nm), with coarse spectral channels (20 nm full width at half maximum), with very high dynamic range (>30 dB) and low insertion loss. The channel spacing was chosen to match the bandwidth over which the LFC amplitude changes in unison, in addition to reducing the overall number of channels. This development also leveraged significant advances in commercial lithographic fabrication techniques over the last few decades. Indeed, SiN fabrication via multi-project wafer runs (fabrication runs where customers pay for only a portion of a wafer), are now producing low loss (<0.5 dB/cm and 0.2 dB/cm more typically at 1550 nm propagation loss) waveguides and devices that are highly reproducible, which was ideal for rapid prototyping of concepts like this. Broadband functional photonics with these characteristics may also find application in biosensing, replacing or complimenting fibers, tapers and Bragg gratings~\cite{Leal2020,Caucheteur:22,Wang:21}.

\begin{figure}[htbp]
\centering\includegraphics[width = \linewidth]{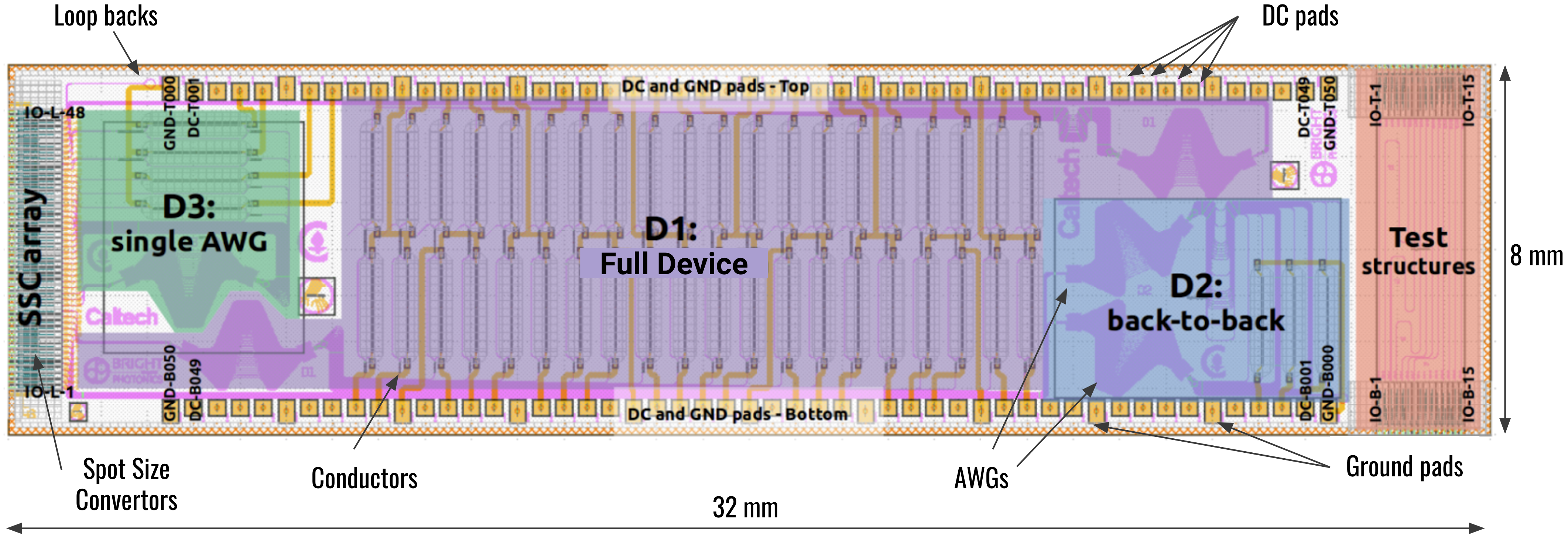}
\caption{Layout of the PIC with the test devices with key elements labelled. Waveguides are in pink and electrodes and pads in yellow. \label{fig:PIClayout}}
\end{figure}

To demonstrate the concept, a prototype device was designed using Nazca Design (\url{https://nazca-design.org/}) with the elements outlined above. In addition to the device, several test structures and circuit elements were also designed to individually characterize the performance of each component. The Photonic Integrated Circuit (PIC) layout is shown in Fig.~\ref{fig:PIClayout}. The image provides an overview with the location and name of various devices and highlights some of the basic elements.

The key elements of the PIC are as follows:
\begin{itemize}
    \item Loop back waveguides used to measure the propagation and coupling losses of the SiN waveguides.
    \item A single AWG used to characterize the properties of a standalone AWG (labelled D3).
    \item A device with back-to-back AWGs (labelled D2) used to measure the spectral reconstruction properties of the AWGs, the relative path length matching of the arms and to see how the spectrum can be modulated in a few channels (1430, 1590, 1770 nm).
    \item A full BAPSS device which consisted of back-to-back AWGs with power and phase modulators in each MZI arm to test the overall concept (labelled D1).
\end{itemize}

\subsection{Arrayed Waveguide Gratings}
\begin{figure}[htbp]
\centering\includegraphics[width = 0.65\linewidth]{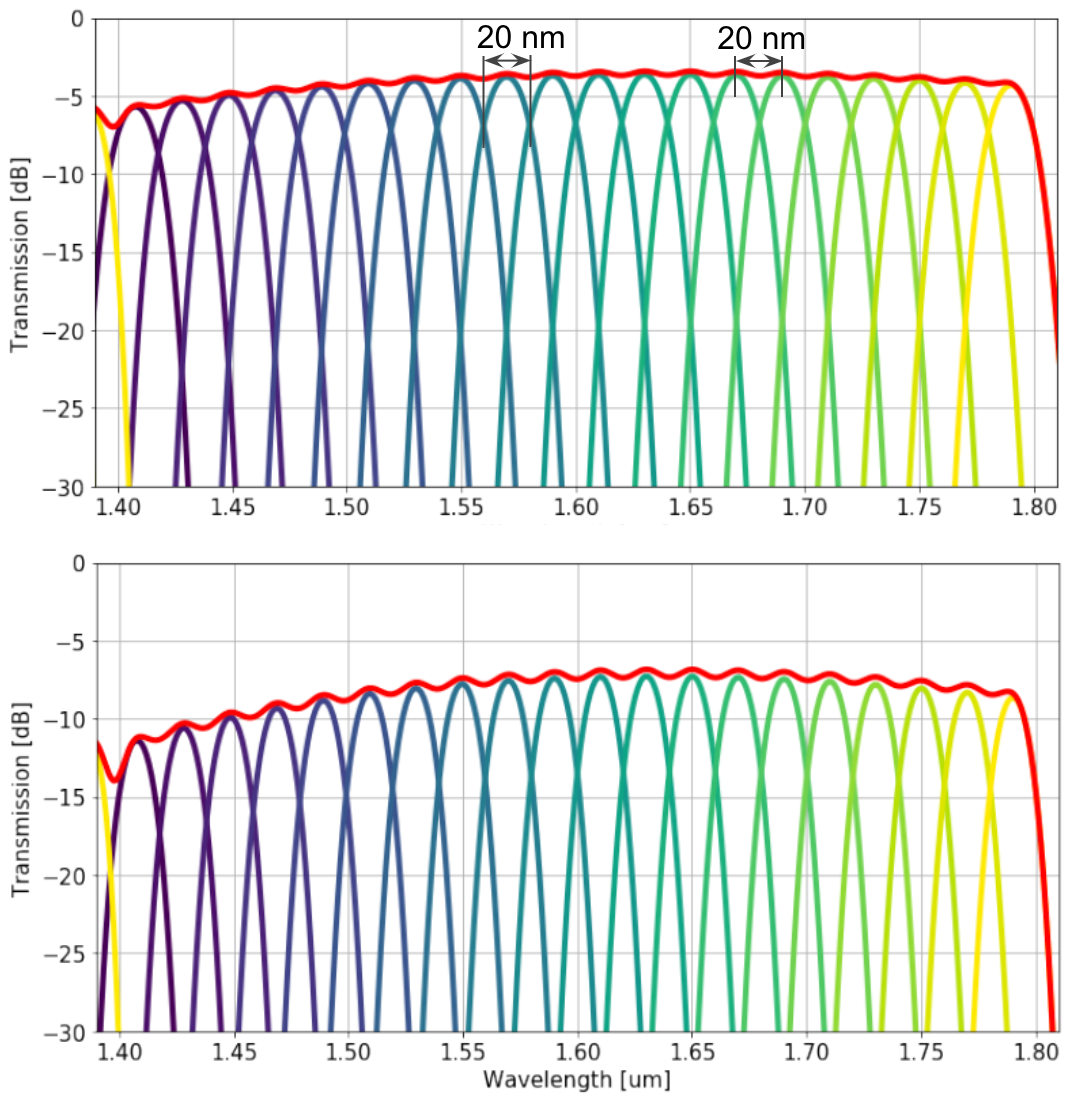}
\caption{Theoretical transmission spectrum of the single pass (top) as well as the back-to-back AWG (bottom). The colors represent each output channel of the AWG. The red line is the cumulative transmission across all channels. \label{fig:AWGspectra}.}
\end{figure}

The theoretical transmission spectra of the AWG as well as the back-to-back AWGs are shown in the top and bottom panels of Fig.~\ref{fig:AWGspectra} respectively. It can be seen that the designed AWG produces Gaussian-shaped spectral channels with $\sim$20 nm wide 3dB bandwidths, spaced $\sim$20 nm apart as required, with no visible side lobes within 25 dB of the peak of each channel.  
The AWG exhibits an insertion loss of $\sim3.5$ dB at 1600 nm which increases to almost 6 dBs at 1400 nm. This roll off in the efficiency of the AWG towards the edges of the free spectral range is a property of the AWG and is related to the far-field illumination of the beams from the waveguides at the output of the array on the output facet  of the free propagation zone. The red curve highlights the cumulative transmission spectrum (i.e. if all channels were summed, as they will be once the light passes through a second stage AWG). The red curve clearly exhibits a ripple which is called spectral dropout and is due to the discretization of the spectrum into waveguides at the output of the device, and some light being lost between them. When two AWGs are used back-to-back, unsurprisingly the losses and spectral drop out double. The theoretical minimum loss of the back-to-back device ranges from 7 dB to 11 dB while the spectral ripple has an amplitude of $\sim1$ dB. These losses are acceptable when it comes to flattening an LFC given the brightness of the source to start with. For other applications that require lower losses, they can be reduced by using lower index platforms like silica-on-silicon, and/or optimizing for a narrower overall wavelength range.

\subsection{Mach-Zehender Interferometers and Multimode Interference Couplers}
\begin{figure}[htbp]
\centering\includegraphics[width = \linewidth]{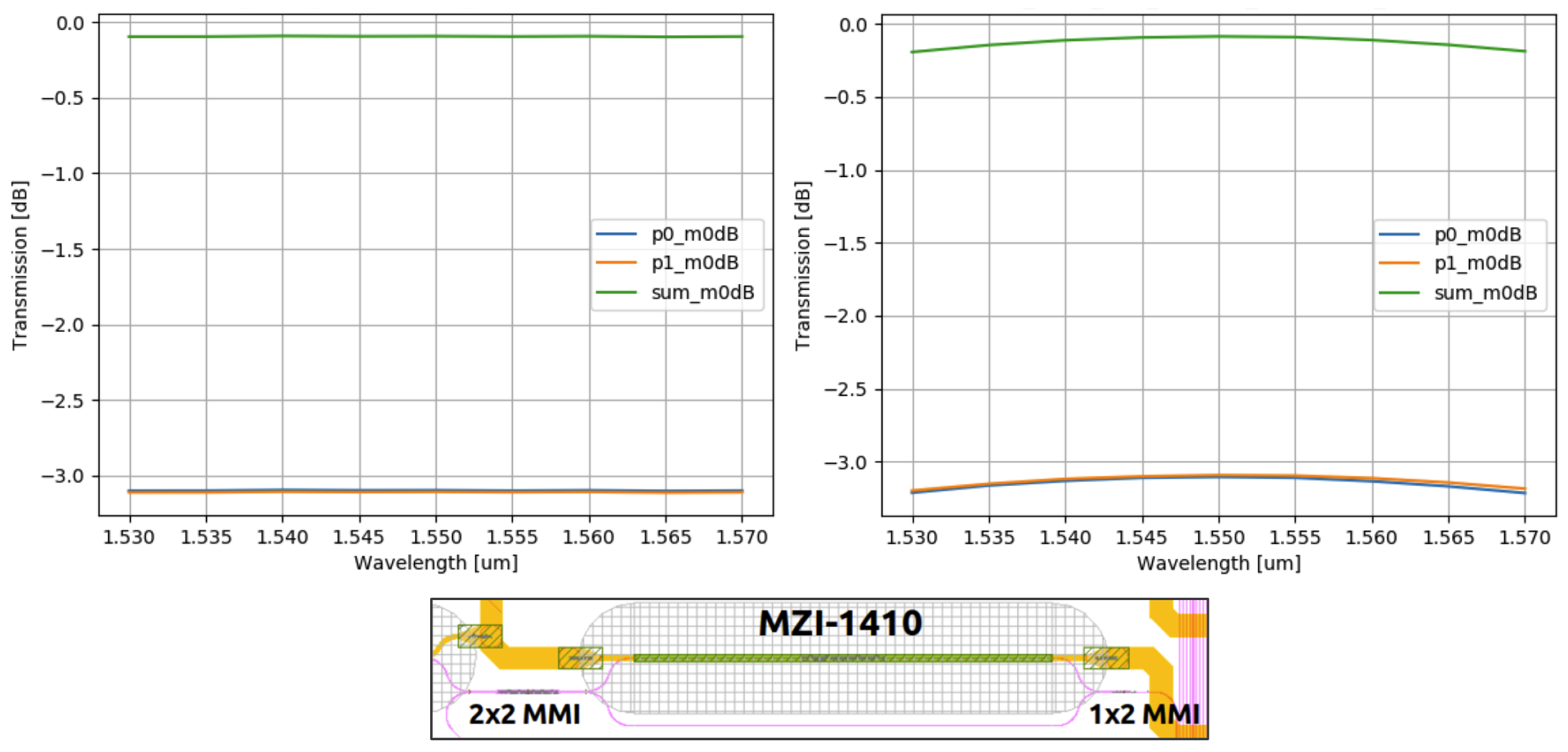}
\caption{(Top) Theoretical transmission spectra of the 1$\times$2 (Left) and the 2$\times$2  (Right) multimode interference couplers. The spectra include the power delivered to each output port along with the total power in the two ports normalized by the input power. At the output of the 2$\times$2 multimode interference coupler the drop port is routed to point away from the rest of the device.  \label{fig:MMIspectra}}
\end{figure}

Between the two AWGs are the MZIs and TOPMs. The MZIs consist of a 1$\times$2 multimode interference coupler, which first splits the light and then a 2$\times$2 multimode interference coupler that recombines the signals. Both multimode interference couplers were designed to split the light 50:50 across the bandwidth of each channel and were therefore optimized for a 40 nm bandwidth to be conservative. The theoretical transmission profiles of the multimode interference couplers, for the 1550 nm channel can be seen in Fig.~\ref{fig:MMIspectra}.  
It can be seen that the splitting ratio for both multimode interference couplers is even amongst the output ports across the entire spectrum. In addition, the multimode interference couplers achieve an insertion loss of $\sim0.1$ dB at the center of the channel and exhibit very little additional losses away from the center wavelength only in the case of the 2$\times$2 variant. Therefore, we can assume that the MZI will introduce an additional 0.2 dB of loss at minimum. 

A 2$\times$2 multimode interference coupler was chosen at the output of the MZI because a 2$\times$1 would result in the rejected light being radiated into the cladding in an uncontrollable way. A 2$\times$2 multimode interference coupler on the other hand allows us to control what happens to the rejected light. We opted to fold the drop port waveguide away from the rest of the structure and terminate it so that the field would radiate away in a controlled manner, minimizing the chance for stray light effects or reflections.

To modulate the MZI and adjust the transmitted power, a TOPM was placed in one arm of the interferometer. Other technologies could be considered for this as well~\cite{Xu_2021}. The static path lengths of the arms of the MZI were set such that they were out of phase by $\sim\pi/3$ radians or $60^{\circ}$. The motivation for this was that then full transmission would occur with only $\pi/3$ radians tuning with the TOPM and the minimum transmission point would occur with a further $\pi$ radians applied. The $\pi/3$ value was chosen to ensure that when manufacturing imperfections were taken into account, we would not overshoot the maximum transmission setting when unbiased.

\subsection{Thermo-Optic Phase Modulators}
Thermo-Optic Phase Modulators follow each of the MZIs to adjust the relative phase of the spectral channels so the spectrum can be recombined as desired. The phase shift is achieved by applying heat to the waveguide region, which modifies the local index via the thermo-optic effect. The phase shift will be different for different wavelengths. However, over the narrow bands of the channels of the AWG ($\sim1\%$ fractional bandwidth channels), the phase shifts are effectively achromatic. 

The TOPMs consist of a chromium electrode deposited onto the top cladding layer above a given waveguide that has a current driven through it. To localize heating to where it is desired the electrodes are made to be wide while routing across the PIC and only narrowed over the top of the waveguide, which increases the resistance (to 500-to 600 ohms) and creates localized heating. The heated regions were 1.9 mm long, which allowed for many $\pi$ radians of phase adjustment. The PIC was laid out in such a way as to keep thermal cross-coupling between any two waveguides to <17 dB. The TOPMs are extremely fast with a switching time of the order of 1 ms as reported by the fabrication foundry used, LioniX International.


\section{Experimental procedures}\label{sec:experiments}
The circuit was fabricated by LioniX International as part of multi-project wafer run $\#$21. The SiN multi-project wafer offering consists of a dual stripe waveguide geometry embedded in a silica cladding. For details about the waveguide geometry refer to the multi-project wafer overview manual (\url{https://www.lionix-international.com/photonics/mpw-services/}).
The waveguides were optimized to allow for optimal guiding around 1550 nm with bend radii as tight as 100 $\mu$m without substantial losses. To improve coupling to optical fibers spot size converters were used at the edges of the chip. These consisted of tapers which would expand 
the mode to $\sim$10 $\mu$m around 1550 nm.

\begin{figure}[htbp]
\centering\includegraphics[width = \linewidth]{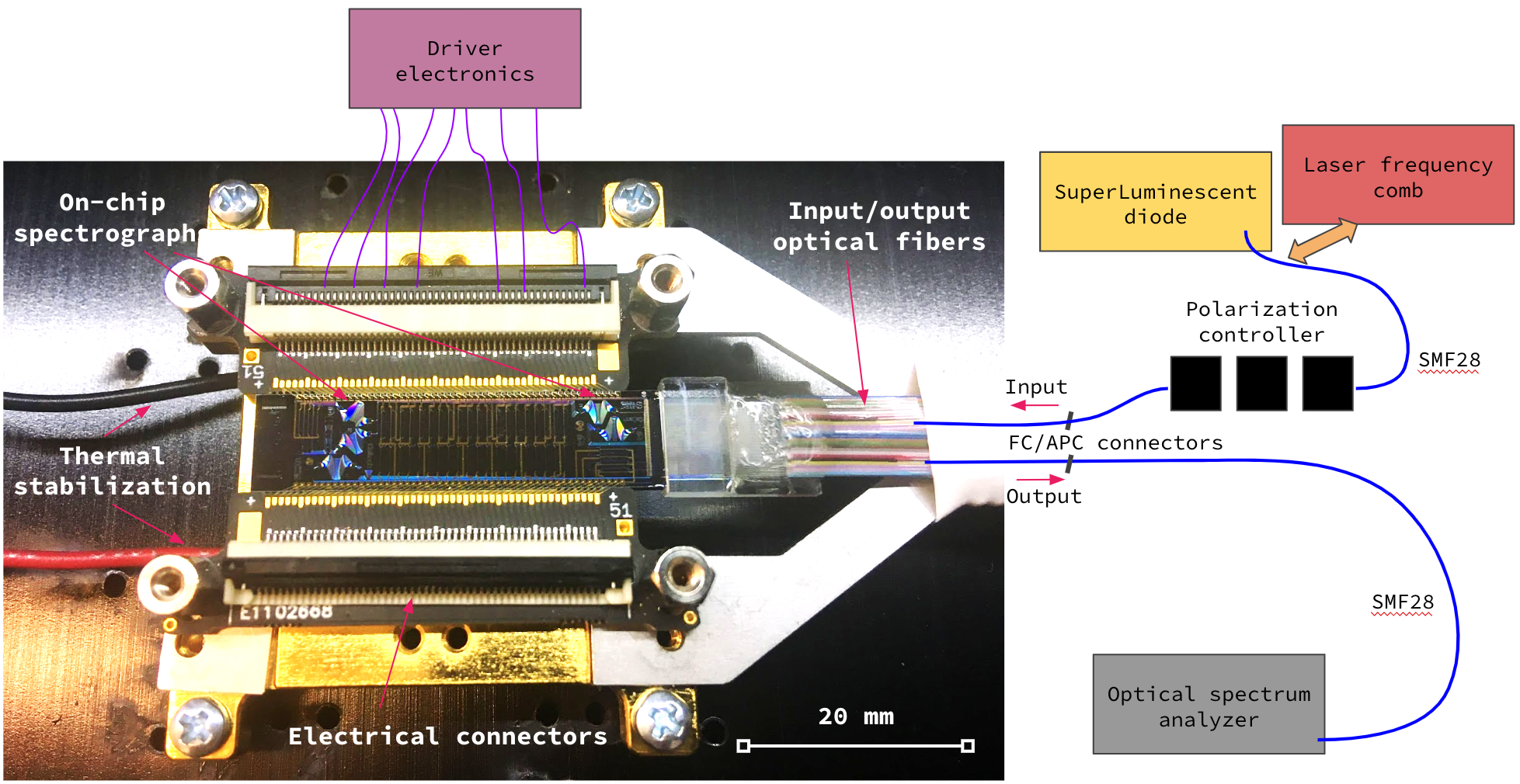}
\caption{Experimental setup for characterizing the broadband all-photonic spectrum shaper. The light source for the tests was either a super-luminescent diode or an LFC. A paddle wheel polarization controller was used to align the polarization of the source with the TE mode of the waveguides. Light was injected via a v-groove array. The fully packaged PIC is shown. The electrodes on the circuit (chromium plated) as well as the AWGs can easily be seen in the photo. The output was routed to an OSA.   \label{fig:setup}}
\end{figure}

After the devices were fabricated they were diced, polished and packaged to simplify testing. This included bonding a 48-fiber vgroove array to one face of the chip. The fiber used for the vgroove was SMF28. All inputs and outputs to and from the various devices were accessed through this v-groove. The circuit was mounted onto a PCB and the DC and ground lines connected so they could be accessed via ribbon cables from the top and bottom of the device. The entire assembly was mounted on a sub-mount that included a themo-electric cooler (which was not utilized for these experiments). An image of the packaged device is shown in Fig.~\ref{fig:setup}. 

To test the devices the setup shown in Fig.~\ref{fig:setup} was used. Initial testing was conducted by injecting light from a super-luminescent diode (Thorlabs, S5FC1550P-A2). Since the fibers bonded to the device under test were SMF28, we used a paddle wheel polarization controller (Thorlabs, FPC561) to orient the polarized signal of the source with the TE mode of the waveguides on the chip, as the waveguides were highly polarization dependent. Note, polarization controllers are chromatic devices which means the polarization could only really be optimized for one wavelength, in our case around 1550 nm, and at wavelengths >50 nm away from this wavelength, there will be polarization cross-talk. The output was connected directly to an optical spectrum analyzer (OSA, Thorlabs, OSA202C).

To test each device, the output of the polarization controller was connected to a given device, the output of that device to the OSA, and then the polarization controller was adjusted to maximize the signal on the OSA. This ensured that the polarization of the light source was aligned with the TE mode of the waveguides, which was most efficient. 

To control the TOPMs we used a multi-channel controller. Since the full device consisted of 20 MZIs and 20 TOPMs, the driver had to support no less than 40 active channels simultaneously. To achieve a maximum of $3\pi$ phase shifts, we required a driver capable of up to 20 V and 50 mA per line. For this reason we used a multi-channel (120) driver from Nicslab (XPOW-120AX-CCvCV-U). A linear power supply was used to power the multi-channel controller and a computer to operate it. To connect the relevant pins of the controller to the PCB and hence the device under test, we used an electrical breadboard (Nicslab, M6 multiconnector). 

Once testing with the super-luminescent diode was completed, we undertook tests with an LFC. However, the comb we had access to was pre-broadened, which meant that it had a triangular shaped spectral profile, not representative of full broadened combs, which are typically flatter, and was only several hundred nanometers wide. Data were acquired with the OSA slit width set to 2 nm resolution, where the lines of the comb were not separated as well as with 0.05 nm resolution where they were.   

The super-luminescent diode had sufficient flux to test the device down to 1400 nm and up to 1650 nm (see inset of bottom panel of Fig.~\ref{fig:AWGresponse} for the profile of the light source). The OSA limited testing at the long wavelength end to 1700 nm, but there is no reason to suspect the devices perform any differently to the data we present in the 1400 to 1700 nm range.

\section{Results}\label{sec:results}

\begin{figure}[htbp]
\centering\includegraphics[width = \linewidth]{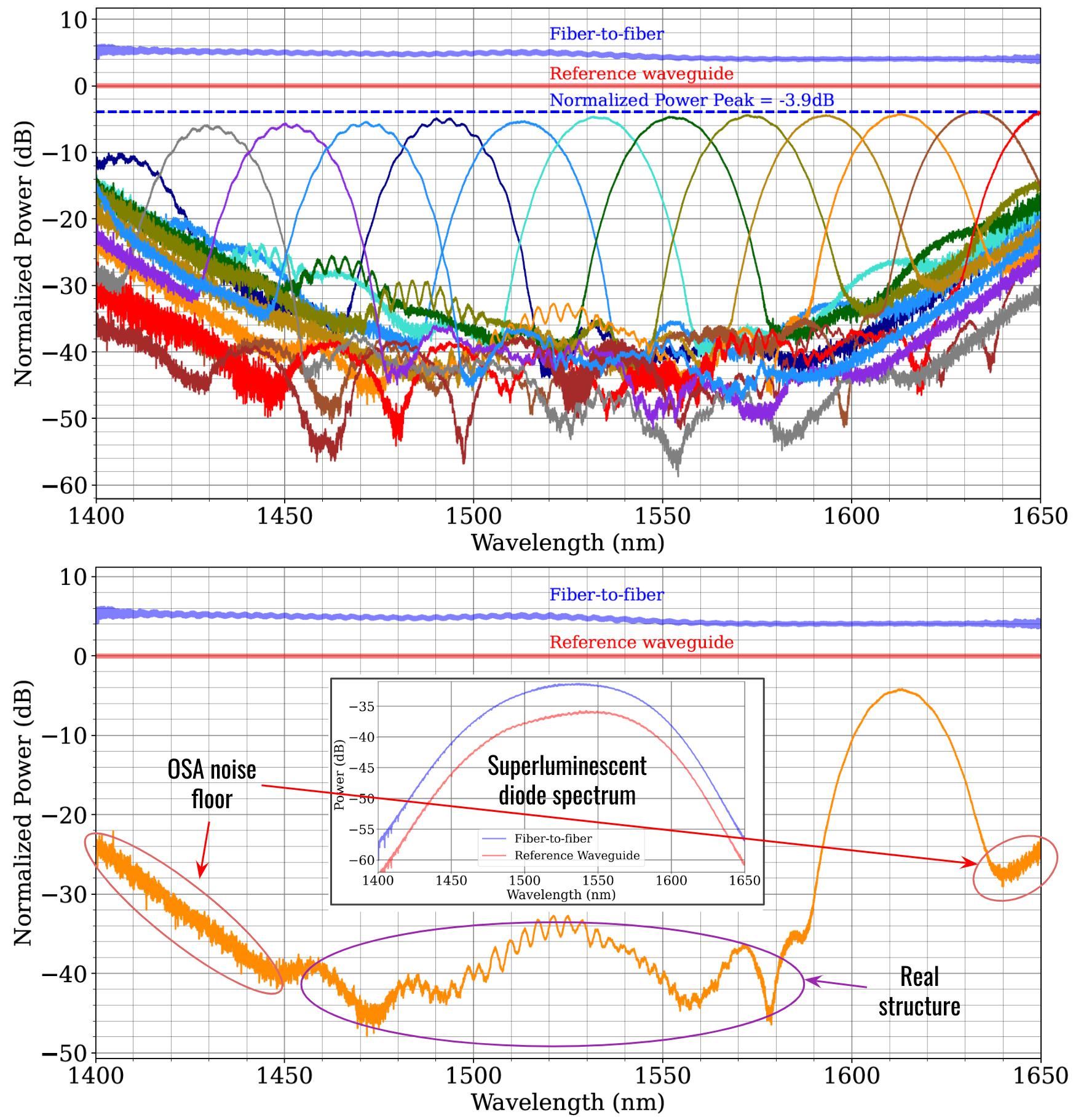}
\caption{(Top) Normalized transmission spectrum of the single AWG. The various spectra represent the response of each spectral channel's output. The data were normalized to the reference waveguide/loop back. This process removes the contribution of coupling losses to/from the chip leaving only the losses of the component/device. The curved base is due to the spectral profile of the light source and represents the noise floor of the measurement. N.B: the noise floors vary because of the varying signal levels of the super-luminescent diode used, which forces the Fourier Transform OSA to switch gain to maximize dynamic range for each measurement. (Bottom) Same plot but only showing the 1610 nm channel so the sub-structure can be more easily inspected. The OSA noise floor is highlighted at the edges of the spectrum. The ripples in the center are real structure indicating there is sub-structure for that spectral channel but at 30+dB below the peak of the channel. (Inset) Spectral profile of the super-luminescent diode light source.   \label{fig:AWGresponse}}
\end{figure}

Using the methods described in the previous section, the spectral response of the single AWG was first probed and the results are shown in Fig.~\ref{fig:AWGresponse}. To generate the spectrum shown, the super-luminescent diode was injected into the input waveguide, and an output waveguide was connected to the OSA, and the response function recorded. We then stepped through all output waveguides systematically recording the response functions. All spectra were normalized to the loop back or reference waveguide, to remove the effect of coupling to and from the device to isolate the losses of the component in question. Since the super-luminescent diode has a bell shaped spectral profile, normalizing all spectra to that corrects for the shape, but imposes a parabolic-like noise floor, as seen in Fig.~\ref{fig:AWGresponse}.

The figure shows cleanly defined Gaussian shaped transmission functions for each channel, which have 3 dB bandwidths of $\sim$20 nm and are spaced by $\sim$ 20 nm, consistent with the design requirements. The channel crossing-points (where the power in neighboring channels is equal), is consistent with the 3 dB bandwidth as designed. The side lobes are >20 dB below the channel peaks as seen in the bottom panel, indicating very low levels of manufacturing errors in the AWGs~\cite{gatkine2021-PCP}. The lowest loss of -3.9 dB is achieved between 1600 to 1650 nm. This value is consistent with the theoretical value (-3.5 dB). The losses of the AWG increase below 1530 nm to about -6 dB at 1430 nm, also consistent with expectation.  

Also shown in the Fig.~\ref{fig:AWGresponse} is the fiber-to-fiber throughput (blue trace). This was taken by connecting the output of the polarization controller directly to the OSA. This removes the entire PIC and by comparing the blue line to the pink line for the reference waveguide we can infer the losses from coupling from fibers in the array to and from the SiN waveguides. The minimum losses are seen around 1600 to 1650 nm and are as low as 4 dB. The loss profile is relatively flat but slowly increases to about 5 dB at 1400 nm. Since the propagation and bend losses in SiN waveguides are very low, and the reference guide length is short, we can assume that the majority of these losses are attributed to coupling to and from the chip. This implies a fiber/waveguide coupling loss of 2 to 2.25 dB per facet over the 1400-1650 nm spectral range. With further effort in engineering the spot size converters we believe this can be reduced. 

\begin{figure}[htbp]
\centering\includegraphics[width = \linewidth]{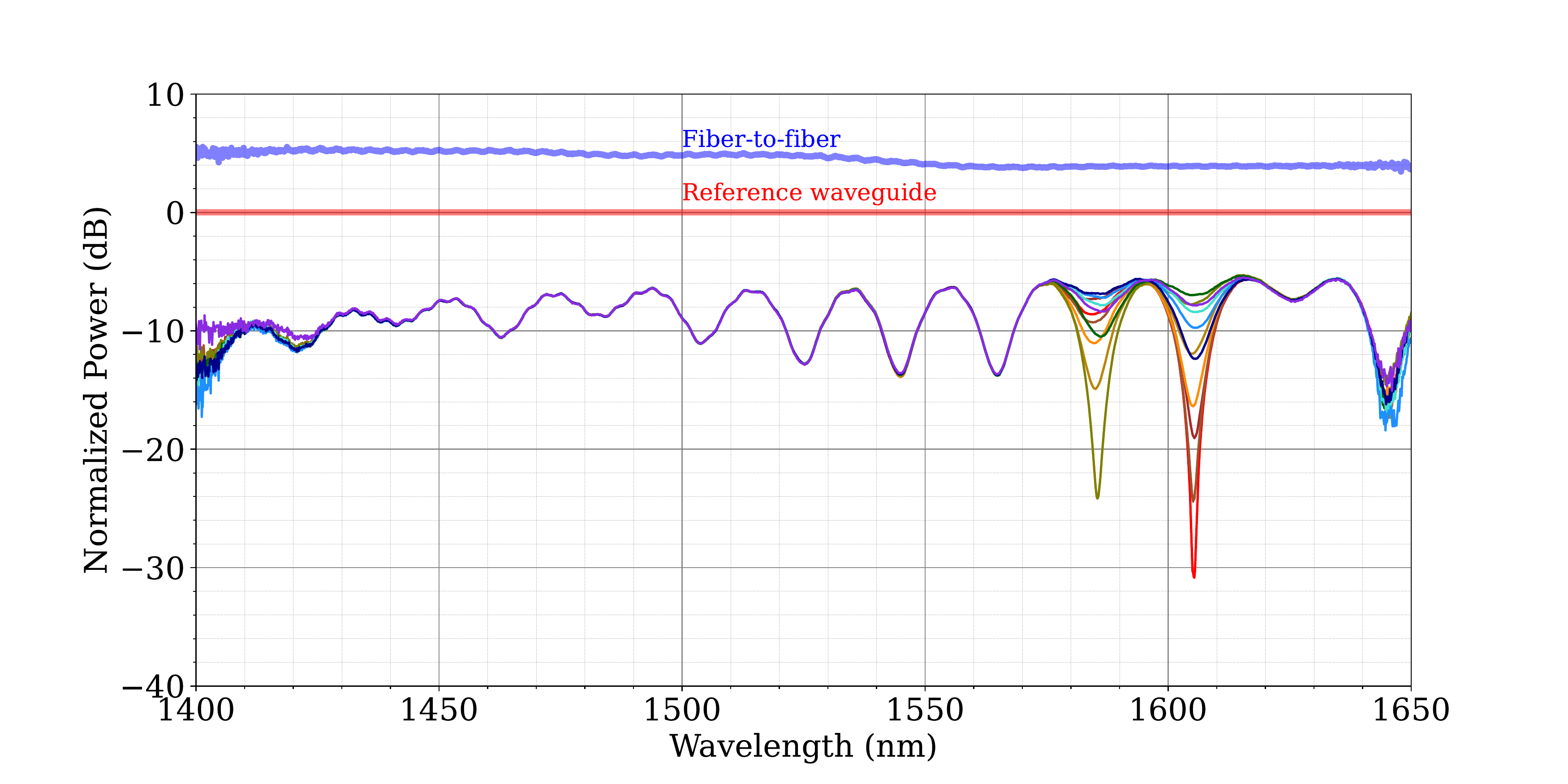}
\caption{Spectral response of the back-to-back AWG device. The 1590 nm channel TOPM has been modulated to show how it can be used to control the amplitude between neighboring channels. The data were normalized to the reference waveguide/loop back. This process removes the contribution of coupling losses to/from the chip leaving only the losses of the component/device.  \label{fig:BBAWGresponse}}
\end{figure}

Figure~\ref{fig:BBAWGresponse} shows the transmission function of the back-to-back AWG device. The spectrum was once again normalized to the reference waveguide. The minimum loss (-6 dB) occurs between 1600 and 1650 nm, which is consistent with simulation to within 1 dB. The losses at the shorter wavelengths increase to -10 dB at 1410 nm, again consistent to within 1 dB of simulations. 

The ripple in the spectrum, due to spectral dropout from the AWGs is evident. The amplitude of the ripple varies between 2 dBs, that which is expected from the models, up to 7 to 8 dBs. This larger than expected dropout is due to the relative phasing of neighboring channels at the time the spectrum is recombined at the second AWG. This means that the static path length compensation implemented in the chip through waveguide routing was imperfect, and the TOPMs had to be used to compensate for the imperfection. 

Figure~\ref{fig:BBAWGresponse} also shows the 1590 nm TOPM being modulated. It can be seen that two regions of the spectrum, on either side of 1590 nm are modulated. This is because only the spectral regions of overlap between neighboring channels are affected when you modulate the phase in the channel. This ``fang'' like spectral response is characteristic of the effect of TOPM modulation in this architecture. It can also be seen that the modulation is unbalanced, i.e. it goes up in one fang while going down in the other (look at the red trace in both fangs to see this), which is a result of the fact the three spectral channels (1570, 1590 and 1610 nm) are not in phase. Once phase matching has been achieved, the fang-like spectral response will modulate both fangs symmetrically. The amplitude of the modulation arising from the TOPMs is >20 dBs.

\begin{figure}[htbp]
\centering\includegraphics[width = 0.9\linewidth]{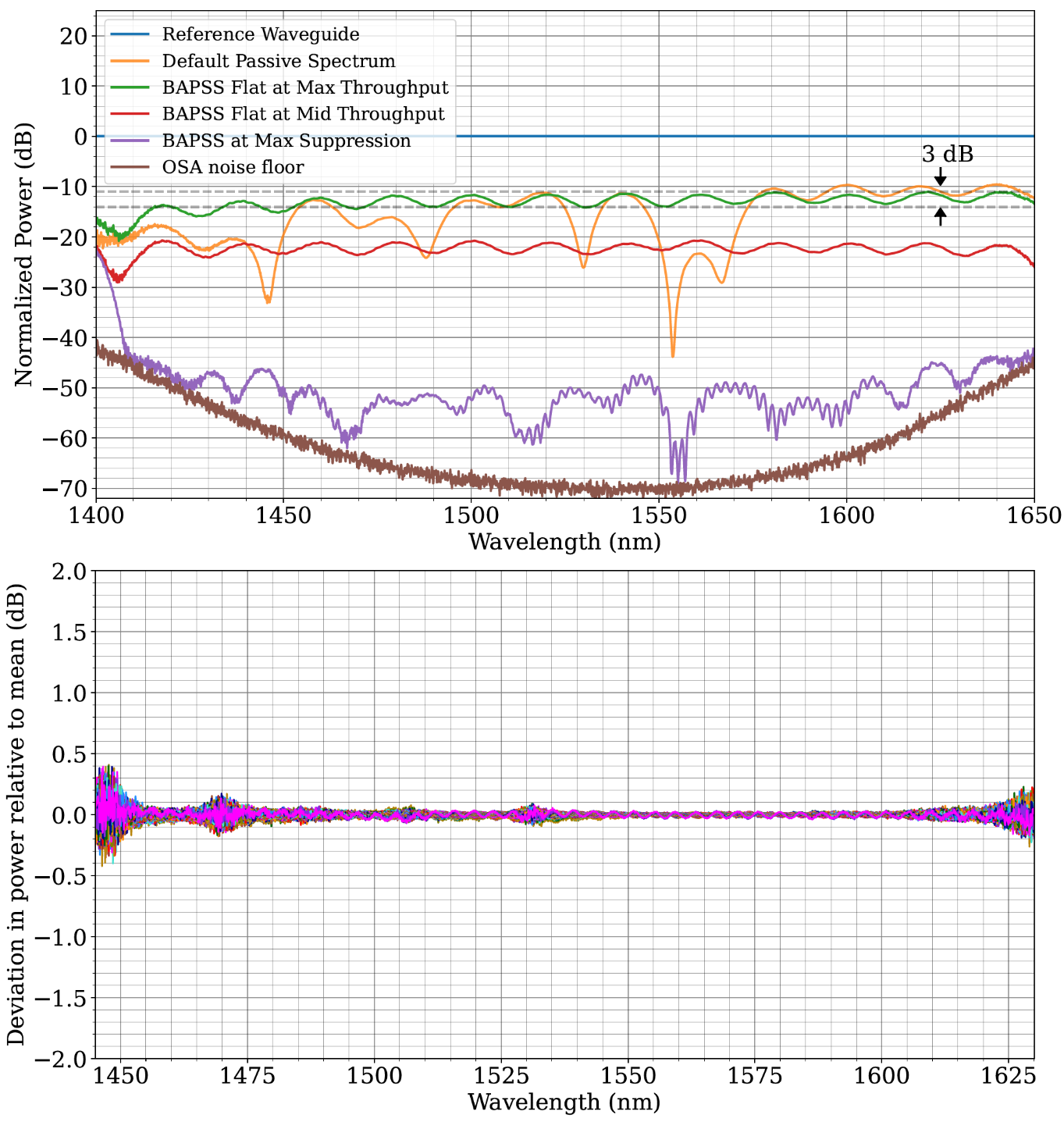}
\caption{(Top) Spectral response of the full device. The orange trace shows the transmission of the device with no power applied. The green is when the MZIs and TOPMS are adjusted for maximum throughput while maintaining flatness. The purple trace is when they are adjusted for minimum throughput, and the red for an arbitrary level in the middle of the range. The data were normalized to the reference waveguide/loop back (blue line at 0 dB). This process removes the contribution of coupling losses to/from the chip leaving only the losses of the component/device. (Bottom) Stability of the spectrum over a 12 hr period with 5 minute cadence. The maximum throughput flat state was used and the figure shows all spectra subtracted from the mean spectrum over that time series. \label{fig:fulldevice}}
\end{figure}

The first two devices have validated that the key building-block components (AWGs, path length matched waveguides) match their expected performance. The next step was to test the entire spectral shaping device. Light was injected into the input of the full BAPSS device and the output routed to the OSA. The results are shown in Fig.~\ref{fig:fulldevice}. Spectra were taken with the MZI's and TOPM's turned off (orange) as well as turned on and adjusted for maximum throughput while maintaining flatness (green), minimum throughput (purple) and some arbitrary flat level in between (red) and are shown in the top panel. 
The minimum losses occur between 1600 to 1650 nm, and are as low as 9.5 dBs. The minimum theoretical losses of the AWG and MZI amount to 7.2 to 7.5 dB. In addition, the approximate path length through the device is 6 to 7 cm and therefore the propagation losses accumulate and add a further 1 to 3 dB which accounts for the remainder of the loss measured. The spectral ripple on the various flat states (green and red) have an amplitude of 2 to 3 dB, which is consistent with what we would expect from the back-to-back AWG. The approximate difference between the maximum and minimum throughput states is 38 dB. The short and long wavelength ends of the minimum throughput spectrum (purple) were impacted by the noise floor of the OSA (brown trace). 

The stability of the full BAPSS device was tested. The BAPSS was set for maximum throughput while maintaining a high degree of flatness (green trace), and the OSA set to record data every 5 minutes for nearly 12 hrs. Note, that the TEC was not used to stabilize the PIC in these experiments, but the lab is stabilized by an air conditioning system to within $\pm0.5^{\circ}$~C. After the experiment was setup, we departed the lab and waited 15 minutes before starting to acquire data to allow the setup to reach thermal equilibrium. The results are shown in the bottom panel of Fig.~\ref{fig:fulldevice}. The data are shown for the region of the spectrum where there was good signal-to-noise on the OSA (1450 to 1625 nm), to not conflate instabilities brought about by the OSA with the BAPSS. It can be seen that traces generated over the $\sim12$ hr period are highly overlapping 
with a scatter of <$\pm0.1$ dB, demonstrating an extraordinary level of stability. 

\begin{figure}[htbp]
\centering\includegraphics[width = \linewidth]{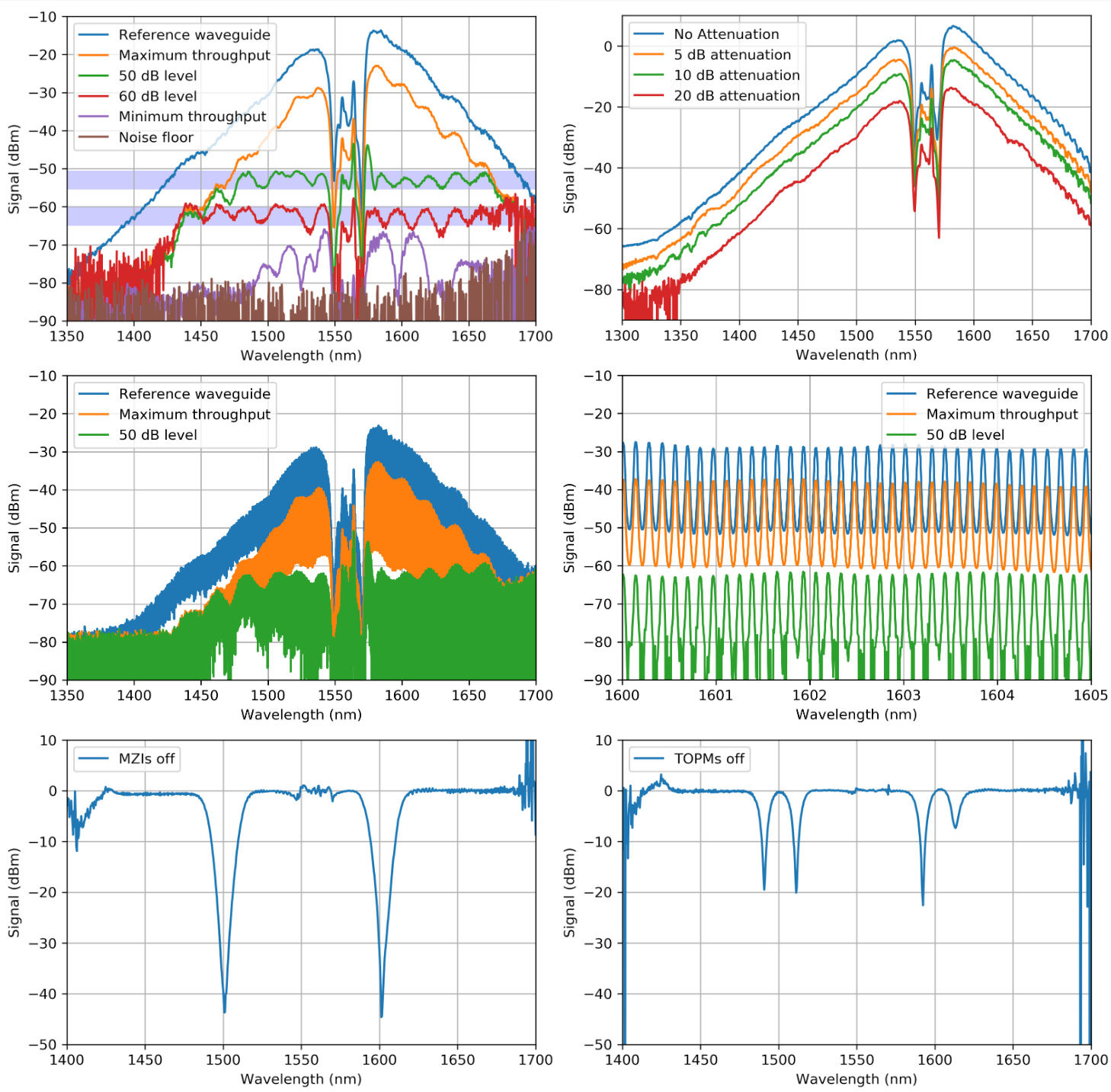}
\caption{Spectral response of the full device with an LFC. (Top left) LFC spectrum through the reference waveguide (blue), and through the BAPSS device with maximum throughput (orange), minimum throughput (purple), and two arbitrary flat levels (green and red). The OSA noise floor is shown in brown. (Top right) the LFC spectrum taken through the reference waveguide with and without optical attenuation. Both plots were acquired with an OSA slit resolution of 2 nm. (Middle row) Shows the same spectrum as the top left plot, but with an OSA resolution of 0.05 nm. The full spectrum is shown on the left and the zoom in of a 2 nm region around 1600 nm is shown on the right. The comb lines are separable in the right plot. (Bottom left) The difference spectrum between the maximum throughput trace with the LFC and with two MZI channels completely turned off. (Bottom right) The difference spectrum between the maximum throughput trace with the LFC and with two TOPMs adjusted to reduce the light passing though the system. Both plots in the bottom row were acquired with an OSA slit resolution of 2 nm. \label{fig:LFCtests}}
\end{figure}

The next set of tests involved the LFC. Figure~\ref{fig:LFCtests} shows the results. The top left panel shows the LFC spectrum through the reference waveguide (blue) as well as the BAPSS device (orange, green, red and purple). It can be seen that the LFC spectrum is triangular in shape on a logarithmic scale, due to the pre-broadened nature of the output of the LFC as outlined in Section~\ref{sec:experiments} above. There is a large dip in the region around 1560 nm because a fiber Bragg grating (FBG) was used to suppress the pump region. The grating was not optimized for the spectral profile or the amplitude of the pump for these tests and so unfortunately offered more suppression over a larger bandwidth than necessary to flatten the spectrum in that region. 

The orange trace shows the same spectrum through the BAPSS device, with the device set to maximum throughput. The green and red traces show the spectrum flattened at the -50 and -60 dBm levels. The flattening was done manually by eye. The blue shaded boxes are 5 dB high, showing that we managed to flatten the spectrum to within 5 dBs, across hundreds of nanometers of range as was the case with the super-luminescent diode tests. The only exception to this is the region around the pump, which has a series of sharp spectral features too narrow for the device to control with the current channel spacing's used. Finally, the purple trace shows how much of the LFC spectrum could be suppressed if we put the device into the minimum transmission state. The brown trace is the noise floor of the OSA indicating that in large sections of the spectrum we are limited by the OSA.  
To re-enforce the capabilities of the MZIs and TOPMs, we show some difference spectra in the bottom row of the figure. These spectra represent the difference between the maximum throughput state and the maximum throughput state with several MZI's (Bottom left) and TOPMs (Bottom right) adjusted for minimum transmission. Once again the MZIs easily exceed 40 dBs of suppression and the TOPMs reach the 20 dB level, consistent with the tests conducted with the super-luminsecent diode. 

Data were also acquired with OSA resolutions of 0.05 nm where the comb lines were separated, as shown in the middle panels of Fig.~\ref{fig:LFCtests}. The panel on the left looks similar to the low resolution scan in the top left panel, except that the traces are no longer lines, but rather shaded regions. This is because in the high resolution mode the modulation in the spectrum due to individual comb lines can be seen. The panel on the right shows the same spectrum but zoomed into only a 2 nm segment around 1600 nm. The lines are clearly separable in this plot. The results at high resolution were deemed to be consistent with the lower resolution data presented here. 

Finally, we tested the power handling capabilities of the device. To ensure we did not damage the device, we connected the LFC to the reference waveguide and took a spectrum. We then replaced the 20 dB attenuator, used in all LFC experiemnts till this point, with a 10 and 5 dB version and finally removed the attenuator altogether. The spectra are shown in the top right panel of Fig.~\ref{fig:LFCtests}. The spectra are largely identical, and are simply offset vertically by the amount corresponding to the attenuation factor. We measured the total integrated power injected into the device with the 20 dB attenuator to be 3 to 4 mW. As we could not directly measure the power with no attenuator, we can infer it would be a 20 dB brighter signal that is achromatic (all traces are parallel to one another) and therefore assume that in the full power mode, 300 to 400 mW of power were being injected into the full device. 

After confirming there were no deleterious effects with these levels of power on the reference waveguide, we connected the LFC to the BAPSS device without any attenuation and conducted some basic tests. The BAPSS device performed identically to how it had with reduced power and is therefore capable of working with power levels of typical LFCs.

\section{Discussion}\label{sec:discussion}
The results above demonstrate that the concept outlined works extremely well for spectral shaping/flattening. Specifically, the device was shown to operate across 250 nm of spectral range, the largest demonstrated from such a layout, with losses consistent with theory. The MZIs were capable of modulating a single spectral channel by 40 dBs, which indicates that the amplitude in the arms of the interferometer is very well balanced, ruling out large manufacturing imperfections. In addition, this modulation range exceeds nearly all bulk optic flatterers which rely on spatial light modulators, which are typically limited to 20 to 30 dBs only. 

The BAPSS flattened a pre-broadened LFC spectrum to a 5 dB amplitude range despite using a chromatic polarization controller. In addition, the LFC polarization was filtered with an off-the-shelf inline polarization filter (Thorlabs, PFC1550A), before injecting it into the BAPSS device, which was also not designed for broad bandwidths. Therefore, the polarization state of the LFC spectrum may not have been perfectly linear and aligned with the TE axis of the waveguides at wavelengths >50 nm from the optimized wavelength (1550 nm) but have no estimate of how imperfect it was. By using bulk optic polarizers to clean up the output of the LFC and PM fibers to deliver the light, the flattening performance should be improved over broader bandwidths.  

The TOPMs offer an extra layer of control that is not present in bulk optic flatteners: the relative phasing between the channels. This produces a distinct ``fang-like" spectral response, that is helpful for understanding the relative phase between neighboring channels.  

The circuit demonstrated here is actually designed for two purposes: static compensation of the native LFC spectrum as well as for tracking the dynamic aspects. In this work we have mostly focused on compensating for the static profile of the spectrum. It should be noted that around the pump region of an LFC there can be sharp features (as seen in Fig.~\ref{fig:LFCtests}), which would require a photonic device with much narrower channels. This can be done, but more channels means more electrodes and increases the complexity of the circuit, which will eventually hit a limit imposed by one of several factors including the size of the reticle. Recently we have designed such a device which occupies a PIC of $32\times16$~mm$^{2}$, has channels with bandwidths ranging from 2 to 20 nm, and supports 39 channels across 1400 to 1800 nm. Going beyond this size and number of channels is possible with customization in several areas: custom lithographic runs with larger reticles and custom PCBs and electronics drivers to support more channels. Scaling to high channel counts (100's to 1000's) is neither practical nor possible and in this regime spatial light modulator based bulk optic flatteners are the more appropriate option. Regardless, the ideal thing to do is to use custom FBGs to compensate for the sharp static aspects of the spectrum, offloading that from the flattener, leaving the flattener to work on the broader parts of the spectrum. If however one wishes to do all of this in the flattener itself, it is possible to use a cascaded AWG approach whereby certain low resolution spectral channels are sent to higher resolution second stage AWGs that provide the sampling needed, but only in a limited number of channels. This is currently being explored and will be focus of future work. 

Although we did not demonstrate the ability of the device to track dynamic changes to the spectrum, the device is more than capable to correct the slow changes expected in an LFC spectrum: with a modulation amplitude of up to 40 dBs and thermal response time as stated by the vendor (and confirmed in a recent publication~\cite{cvetojevic2022-SCS}) of 1 ms it will be sufficient to correct for the slow evolving, low amplitude changes typical of LFC spectra. How to implement the control law for closed loop operation will require some consideration as well, especially since the TOPMs have a complex spectral response. One possible avenue is to use the TOPMs to phase all channels in the array, and then keep them fixed and only drive the MZIs. This reduces the degrees of freedom dramatically and simplifies the problem to that akin to SLM based versions. To generate the signal used to drive the tracking loop however, one would ideally substitute the OSA with the actual spectrograph used in for example astronomical observations. This has the advantage of allowing the BAPSS to be used to compensate for any other wavelength dependent loss differences between the BAPSS and the spectrograph. However, this is impractical because 1) the timescales of changes are faster than typical readout times used for astronomical spectra acquisition and 2) the data would need to be extracted in real time, which is not feasible. Therefore, the baseline approach could be to split off a portion of the signal at the output of the BAPSS and send it to a small dedicated portable lab spectrometer, which are common place in many labs. Another alternative would be to integrate photodetectors directly onto the PIC, or use a flip chip design, where the photodetectors could be used to sense either the rejected signal from the $2\times2$ multimode interference couplers or some signal tapped after the TOPM in each channel. Such an approach would offer a miniature and portable solution to spectral shaping.  

The fact the device performed so close to the specification on our first attempt to fabricate it is a testament to how SiN multi-project wafer offerings have advanced and are now capable of producing very high quality PICs. This is a critical demonstration on the path towards rapid prototype photonics and highlights that high performance spectral shapers/flatteners on a chip can be readily fabricated for reasonable costs. Mass production can of course reduce this cost further.  

Astronomical spectroscopy typically requires large bandwidths. We demonstrated successful operation of the BAPSS over a single band (1490 to 1800 nm) in this work. It may be possible to re-engineer the device to increase the bandwidth to capture several bands and/or split the light at the output of the LFC and send it to several devices each optimized for operation in a different band. This is another avenue that can be explored in future.


\section{Summary}
We have demonstrated an all-photonic device capable of shaping and specifically flattening a spectrum over several hundred nanometers in the NIR (1400 to 1800 nm). The device is based on standard photonic components including AWGs, MMIs, MZIs and TOPMs and was fabricated on  a SiN chip via the a multi-project wafer scheme. The device demonstrated 40 dB of modulation from the MZIs, and $>20$ dB from the TOPMs, across the 250 nm of spectrum that was tested, indicating that the components in the circuit are of high quality and are defect free. The device flattened a super luminescent diode spectrum to 3 dB, the limit imposed by the components in the circuit, and an LFC to within 5 dBs, from an initial 25 dB+ amplitude range. With improvements in polarization control of the LFC light, the spectrum could be made flatter. This demonstrates the possibility to re-optimize telecommunications components for broadband applications as well as the viability of this circuit layout and fabrication process for flattening LFC spectra.    

Besides calibration source optimization for precision radial velocity measurements, the BAPSS demonstrated here may find applications in optical pulse shaping, dynamic gain flattening and targeted molecular spectroscopy applications and possibly even nulling interferometry for exoplanet detection. Given that SiN waveguides have been demonstrated to operate from 400 nm up to 2300 nm, and AWGs are readily fabricated around 800 nm~\cite{Rank2021-TOC}, its reasonable to expect that this concept can be shifted into other wavebands with some optimization to the devices. A shift into other wavelength regions will extend the reach of this technology.


\begin{backmatter}
\bmsection{Funding}
This work was produced with support from the W.M. Keck Institute for Space Studies. 

\bmsection{Acknowledgments}
This work was produced with support from the Keck Institute for Space Studies. N. Cvetojevic acknowledges the funding from the European Research Council (ERC) under the European Union's Horizon 2020 research and innovation program (grant agreement CoG - 683029). P. Gatkine is supported by NASA through the NASA Hubble Fellowship grant HST-HF2-51478.001-A awarded by the Space Telescope Science Institute, which is operated by the Association of Universities for Research in Astronomy, Inc., for NASA, under contract NAS5-26555. Gatkine also acknowledges the support from David and Ellen Lee Postdoctoral Fellowship at the California Institute of Technology. Some of this research was carried out at the Jet Propulsion Laboratory, California Institute of Technology, under a contract with the National Aeronautics and Space Administration (80NM0018D0004). 

\bmsection{Disclosures}
The authors declare no conflicts of interest.

\bmsection{Data Availability Statement}
Data underlying the results presented in this paper are not publicly available at this time but may be obtained from the authors upon reasonable request.

\end{backmatter}



\bibliography{references}

\end{document}